# OCTOPUS – optical coherence tomography plaque and stent analysis software

Juhwan Lee[1], Justin N. Kim[1], Yazan Gharaibeh[1], Vladislav N. Zimin[2], Luis A. P. Dallan[2], Gabriel T. R. Pereira[2], Armando Vergara-Martel[2], Chaitanya Kolluru[1], Ammar Hoori[1], Hiram G. Bezerra[3], David L. Wilson[1,4,*]

[1] Department of Biomedical Engineering, Case Western Reserve University, Cleveland, OH, 44106, USA
[2] Cardiovascular Imaging Core Laboratory, Harrington Heart and Vascular Institute, University Hospitals Cleveland Medical Center, Cleveland, OH, 44106, USA
[3] Interventional Cardiology Center, Heart and Vascular Institute, University of South Florida, Tampa, FL, 33606, USA
[4] Case Western Reserve University, Department of Radiology, Cleveland, OH, 44106, USA

*Corresponding author: dlw@case.edu
  Telephone number: 216-368-4099, fax: 216-368-4969

**Abstract**
Compared with other imaging modalities, intravascular optical coherence tomography (IVOCT) has significant advantages for guiding percutaneous coronary interventions. To aid IVOCT research studies, we developed the Optical Coherence TOmography PlaqUe and Stent (OCTOPUS) analysis software. To automate image analysis results, the software includes several important algorithmic steps: pre-processing, deep learning plaque segmentation, machine learning identification of stent struts, and registration of pullbacks. Interactive visualization and manual editing of segmentations were included in the software. Quantifications include stent deployment characteristics (e.g., stent strut malapposition), strut level analysis, calcium angle, and calcium thickness measurements. Interactive visualizations include $(x,y)$ anatomical, en face, and longitudinal views with optional overlays. Underlying plaque segmentation algorithm yielded excellent pixel-wise results (86.2% sensitivity and 0.781 F1 score). Using OCTOPUS on 34 new pullbacks, we determined that following automated segmentation, only 13% and 23% of frames needed any manual touch up for detailed lumen and calcification labeling, respectively. Only up to 3.8% of plaque pixels were modified, leading to an average editing time of only 7.5 seconds/frame, an approximately 80% reduction compared to manual analysis. Regarding stent analysis, sensitivity and precision were both greater than 90%, and each strut was successfully classified as either covered or uncovered with high sensitivity (94%) and specificity (90%). We introduced and evaluated the clinical application of a highly automated software package, OCTOPUS, for quantitative plaque and stent analysis in IVOCT images. The software is currently used as an offline tool for research purposes; however, the software's embedded algorithms may also be useful for real-time treatment planning.

# 1    Introduction

Percutaneous coronary intervention (PCI) is the most common revascularization procedure, with an average of 700,000 procedures performed per year in the United States [1]. Major calcifications are of great concern when performing PCI [2], since it can lead to under-expansion and strut malapposition [3,4]. Lack of stent strut coverage has been associated with increased risk of thrombosis and in-stent restenosis [5,6]. Therefore, accurate calcified plaque characterization and stent deployment analysis are essential for PCI evaluation and follow-up.

Intravascular optical coherence tomography (IVOCT) has rapidly evolved in the past decade and has become very useful for interventional cardiology [7,8]. This imaging modality not only provides quantitative measurements of the lumen, but also detects fibrous tissue, lipid tissue, calcium, and macrophage deposition [9–12]. Although IVOCT has many clinical advantages, a single pullback typically generates more than 500 image frames. Complete manual analysis of this large amount of data is time-consuming and labor-intensive, making application of IVOCT to real-time treatment planning challenging. We hypothesized that fully automated analysis of IVOCT images would provide faster and comprehensive assessment of plaque to support treatment decision making. Furthermore, quantitative assessment of plaque and stent deployment would allow longitudinal analysis of the same lesion, which may advance our understanding of lesion progression and the response to PCI treatment.

Our group has proposed several IVOCT image analysis approaches, including plaque characterization and stent strut analysis. We developed an automated method for estimating the optical properties of coronary plaques, including calcified, fibrotic, and lipid atherosclerotic plaques [13]. We also developed machine and deep learning methods to identify fibrolipidic and fibrocalcific A-lines in IVOCT images [14,15]. In other studies [16–19], we successfully implemented a deep learning-based semantic segmentation to characterize coronary plaques, such as lipid and calcification. Recently, we proposed a hybrid learning approach, which combines deep learning and hand-crafted, lumen morphological features for atherosclerotic plaque characterization [20]. We also created machine learning-based approaches to analyze stents in great detail, including individual strut tissue coverage [21–24].

In this paper, we introduce the first version of Optical Coherence TOmography PlaqUe and Stent (OCTOPUS) analysis software, which provides highly automated, comprehensive analysis of coronary plaques in IVOCT images. To the best of our knowledge, this is the first available software for quantitative plaque analysis in IVOCT images. OCTOPUS can be applied to both offline analysis of IVOCT images and real-time treatment planning in clinics, thereby enabling faster, more accurate treatment decision making. In addition, OCTOPUS can process multiple pullbacks at one time, with versatile functions that facilitate more comprehensive analyses. Users can load and analyze up to four pullbacks simultaneously through our specially designed viewers. This feature is particularly useful for comparing baseline and follow-up pullbacks obtained at different time points. Finally, OCTOPUS can be used to advance our understanding of plaque morphology, including automated measurements of lumen and calcification attributes, and can help determine the stent landing zone by providing cross-sectional, longitudinal, and en face images.

# 2    Image Analysis Methods

This section briefly describes the core algorithms included in the OCTOPUS software: pre-processing, stent strut analysis, lumen/calcification segmentation, co-registration between pullbacks, and plaque quantification.

*2.1 Guidewire and corresponding shadow removal*

The guidewire does not provide meaningful information for plaque analysis, as it appears as a bright arc followed immediately by a dark shadow due to its high reflectivity. Therefore, we removed the guidewire and corresponding shadow regions using an accumulated intensity map [25]. Briefly, the intensity map was generated by adding the values of pixels on each A-line, and the summation was normalized. The guidewire typically presents the upper and lower boundaries of the polar ($r,\theta$) domain, which are easily distinguishable from the background. To segment these boundaries, we used dynamic programming on the dark-bright and bright-dark boundaries [25]. Dynamic programming is an algorithmic technique used to solve optimization problems by breaking it down into simpler sub-problems. We defined the guidewire detection as searching the point that maximizes the intensity differences in the polar ($r,\theta$) domain. In our previous study, the detected guidewire boundary well matched the actual position for 9 entire clinical pullbacks [25]. The guidewire and corresponding shadow were removed from further consideration.

*2.2 Lumen segmentation and pixel-shifting*

Robust lumen segmentation is important for calculation of lumen attributes and for later plaque segmentation steps. We used our previously developed deep learning method [17] to segment the lumen boundary in IVOCT images. Briefly, the SegNet semantic segmentation model [27] was used for lumen segmentation with the VGG-16 pre-trained model [28], and the results were post-optimized using the opening morphological operation with a kernel size of 5 to effectively remove small island errors. A total of 2,640 image frames from 34 pullbacks was used for training network. In our previous study [17], we found that deep learning lumen segmentation was superior to our earlier dynamic programming lumen segmentation [25]. After lumen segmentation, each A-line was pixel-shifted to the left in $(r,\theta)$ images, so that the lumen border was at the same horizontal location in the $(r,\theta)$ array. We only considered a depth of 1.5 *mm* (300 pixels) in the *r* direction from lumen border due to limited penetration of the near-infrared signal. Gaussian filtering was applied to reduce noise. The resulting image was used as the input for the deep learning models. Algorithm details are described elsewhere [15,16].

*2.3 Determination of major calcification lesions*

Calcified plaque is defined as a signal-poor area with sharply delineated borders in IVOCT images [29]. We used a 3D convolutional neural network (CNN) [18], which considered 8,231 frames across 68 pullbacks, to classify frames containing major calcified lesions prior to plaque segmentation. The 3D CNN is composed of five convolutional, five maximum pooling, and two fully connected layers. Each convolutional layer included convolutional, batch normalization, and rectified linear unit layers. Convolutional processing was done under a varying number of filters (96, 128, 256, and 324), with a filter size of 5x3x5 pixels and a stride of 1. For the maximum pooling layer, the pool size was set to 2x2x1 pixels. After the initial determination of major calcification, we applied morphological opening and closing operations using a kernel size of 3 to more efficiently detect major calcifications. The proposed 3D CNN model was superior to all other employed 2D models. Algorithm and training details are described in [18].

*2.4 Calcification segmentation*

In frames containing calcified lesions, calcified plaques were segmented using a deep learning approach, which was the most important and challenging task for the OCTOPUS software. Briefly, as reported previously [18], we used a SegNet convolutional neural network classifier consisting of five encoding and five decoding steps, followed by a pixel-wise classification layer. The learning parameters were initialized by transferring weights from a VGG-16 pre-trained network [28], and adaptive moment estimation optimization was used for better training. To effectively reduce classification errors, we implemented a fully connected conditional random field [30] after segmentation. We chose a SegNet model because this model had better performance than other state-of-the-art semantic segmentation models in our dataset [16]. More than 12,000 IVOCT images including 8,231 clinical and 4,320 ex-vivo cadaveric images were used to train the network. The clinical dataset was acquired from 68 patients having 68 pullbacks, and the ex vivo cadaveric dataset was obtained from 4 heavily calcified arteries [18].

*2.5 Comprehensive stent deployment analysis*

Automated stent analysis includes stent strut detection, stent strut classification (covered or uncovered), stent contour estimation, and tissue coverage and malapposition quantification [23]. For stent strut detection, the candidate struts were first detected using image characteristics (bright reflection followed by a dark shadow) [21]. Then, image features were extracted from the strut bloom and shadow, and true stent struts were classified using a bagged decision tree. Stent tissue coverage is defined as tissue overlying the struts >0 $\mu m$ [31] and is an important surrogate biomarker of stent viability [6,32–35]. To examine the presence of tissue coverage on top of the strut, we calculated 21 hand-rafted features from center and side patches [22], and each strut was classified as either covered or uncovered using a support vector machine (SVM). We created the stent contour by interpolating the locations of each strut in $(r, \theta)$ images. Lastly, tissue coverage thickness and malapposition distance were measured from strut center to its closest-distance point on the lumen boundary. The IVOCT data set used to create machine learning model consisted of 7,125 image frames including 39,000 covered and 16,500 uncovered struts.

*2.6 Registration of IVOCT pullbacks*

Co-registration of IVOCT pullbacks will enable many studies, including studies of plaque progression/regression and the effects of plaque characteristics obtained from pre-stent images on stent deployment and neo-atherosclerosis.

We used both manual registration of landmarks (e.g., side branches or stents) and an automated approach suitable for registration within a lesion. In the former case, two identifiable landmarks are marked on the reference and floating pullbacks. The floating pullback is then offset to match the reference. For the automated approach, the IVOCT data acquired in a spiral is unfolded into a continuous 2D ($r$, $\theta$) array, where $\theta$ increases from 0 to $N$x360 deg and $N$ is the number of IVOCT frames. Calcifications are segmented using a deep learning approach to create a 1D plot of maximum calcification thickness. For a pair of IVOCT pullbacks, cross-correlation of two such curves is performed to find the offset, giving a peak correlation, and the offset value is used to register the 2D arrays. The "floating" 2D array is then reassigned to frames that match the reference pullback. A preliminary report describes this method [36]. For cases where we want to register a pre-stent pullback to a pullback with a stent, we apply a generalized adversarial network approach to create a virtual pullback [37] without stent struts and then apply the automated registration method.

*2.7 Plaque quantification*

Quantitative tissue analysis is essential for selecting the optimal treatment approach. The Z-offset was manually calibrated [8] to provide accurate plaque measurements. For better interpretation of plaque distribution, the lumen and calcification attributes were estimated from automated segmentation results. Lumen attributes include the area and maximum/minimum/mean diameters, while calcium attributes consist of maximum angle, maximum thickness, minimum depth, and length (Fig. 1).

# 3 Software description

*3.1 Key functions of OCTOPUS software*

OCTOPUS was developed using MATLAB (R2019b, MathWorks, Inc) on a NVIDIA Geforce TITAN RTX GPU (64 GB RAM) and was compiled to a standalone package with the MATLAB Compiler v9.7. MATLAB Compiler Runtime v9.7 is required to run the software. Figure 2 shows the main graphical user interface (GUI) of OCTOPUS software. OCTOPUS is fully automatically operated through its dedicated GUI. The key functionalities included in the software are described below.

**Input and output.** OCTOPUS supports *.oct image data (Abbott Vascular Inc., Santa Clara, CA, USA) as well as DICOM. The software is initiated by loading an IVOCT pullback using the *Open File* button on the top-right of the GUI (Fig. 2). For convenience, multiple pullbacks can be loaded. The outputs of the software include grayscale images in ($x,y$) and ($r,\theta$), corresponding label images, en face maps, and Excel files summarizing all quantitative assessments for further offline analysis.

**Settings.** After loading IVOCT pullback(s), the user selects *Mode* (Fig. 2), which has three potential values: *Baseline*, *Follow-up*, and *Stent Analysis*. *Baseline* would normally correspond to a pre-stent pullback. It enables plaque characterization and visualization. There are five independent views: ($x,y$) anatomical, three en face ($\theta,z$), and longitudinal ($z$). *Range* allows analysis and visualization of an entire pullback or a specific region of interest (ROI) designated with start and end frames. As shown in Fig. 3, the *Follow-up* mode enables review of multiple pullbacks at the same time in up to four viewers. *Stent Analysis* enables comprehensive stent deployment analysis (Fig. 4).

**Co-registration between pullbacks.** Within the *Follow-up* mode (see Fig. 3), IVOCT pullbacks can be manually or automatically co-registered to the first viewer. For example, if the user manually identifies the matching frame, the corresponding frame is synchronized with the first viewer by clicking the *PB* button under the *Co-registration* menu. In the *Automatic* mode, the software automatically detects the matching frame and co-registers all the loaded pullbacks. If the automatic co-registration is inappropriate, the user can manually modify the matching frame.

**Plaque segmentation.** Plaque segmentation is a key function of OCTOPUS software and is carried out automatically with predefined ranges (either All or ROI) selected in the *Settings* menu (Fig. 2). The current version of OCTOPUS provides lumen and calcium segmentations. The user can select the one of tissue types among all, lumen, and calcification through the *Automatic Segmentation* menu. Plaque characterization is then performed by clicking the *Run* button. The software provides a progress bar during image processing, so that the user can monitor progress status.

**Viewers.** As briefly described earlier, OCTOPUS provides different types of viewers according to the selected run mode. For the *Baseline* mode (Fig. 2), the left viewer displays cross-sectional (*x,y*) IVOCT images and enables manual plaque segmentation and/or editing of automated results. The three viewers on the right present en face maps of angle, thickness, and depth of calcifications from top to bottom, respectively. This helps analysts better understand the plaque distribution and readily determine the stent landing zone. The bottom viewer shows the longitudinal image overlaid with the plaque segmentation in red. The *Follow-up* mode aids users by simultaneously analyzing multiple IVOCT pullbacks using up to four viewers. For example, on the *Follow-up 2* mode, the software provides two cross-sectional viewers (top) and corresponding longitudinal maps (bottom) (Fig. 3). The longitudinal map is changed according to the projection angles (red-green line) in the cross-sectional viewer, thereby helping to clearly identify plaque distributions.

**Image editing.** OCTOPUS provides not only automated plaque characterization, but also an interactive editing function. The *Editing Functions* mainly include the two sub-menus of *LABEL* and *ADD* that allow users to modify existing labels and add new labels, respectively. The labels are modified using the interactive editing tools (e.g., freehand and brush) and are automatically updated and exported. The user can add four classes of labels: lumen, lipid, calcium, and other.

**Measurements.** After plaque segmentation, OCTOPUS provides quantitative measurements of lumen and calcification attributes in the bottom right window (Figs. 2 and 3). All measurements are summarized for each frame and exported as an Excel file to the local directory. Additionally, the user can manually measure the angle and length of a desired region on the cross-sectional viewer and length in *z*-direction on the en face and longitudinal maps. Manual measurements are annotated and saved in the current frame.

*3.2 Manual correction evaluation*

Manual correction analysis was performed on a total of 12,750 IVOCT images from 34 patients with 36 calcified lesions. All images were acquired with a frequency-domain ILUMIEN OCT system (St. Jude Medical Inc., St. Paul, Minnesota, USA). Imaging pullback was done on *Survey* mode with a pullback speed of 36 *mm/s* and pullback length of 75 *mm*. All IVOCT images were initially reviewed by an expert reader. Exclusion criteria were poor image qualities due to luminal blood, unclear lumen, artifact, or reverberation. All images were manually labeled based on consensus of two expert cardiologists from the Cardiovascular Imaging Core Laboratory, Harrington Heart and Vascular Institute, University Hospitals Cleveland Medical Center, Cleveland, OH, USA.

# 4   Results

We first relate some of the underlying algorithms in OCTOPUS, starting with detection and segmentations of calcifications. The OCTOPUS software accurately identified calcified plaques in most situations; an example is shown in Fig. 5. In our previous publication [18], we trained/tested the deep learning networks on a large dataset including 8,231 in vivo clinical images from 68 vessels and 4,320 ex vivo cadaveric images from 4 vessels. The 3D CNN model successfully identified the major calcifications with high sensitivity (97.7%), specificity (87.7%), and F1 score (0.922). A SegNet deep learning model showed reliable segmentation results with 86.2% sensitivity and 0.781 F1 score. An example of stent strut detection is shown in Fig. 6. The algorithms provided reliable stent deployment analysis in post-PCI pullbacks (Fig. 6). On the 80 pullbacks, including 39,000 covered and 16,500 uncovered struts, sensitivity and precision of strut detection were greater than 90% [21]. Each strut was classified as either covered or uncovered using a SVM, giving high sensitivity (94%) and specificity (90%) [22]. We also evaluated strut tissue coverage thickness of covered struts and found better measurements than those made using the commercial product, with the mean error of 2.5±29.0 *μm*. Automated image registration is demonstrated in Figs. 3, 7, and 8. As the automated method is applicable to regions with calcifications, we also created a semi-automated method, which was also found to work well under the direction of an experienced analyst.

OCTOPUS is designed to provide highly automated, accurate analyses with the potential for manual editing. We analyzed 34 pullbacks and 2,723 image frames. We found that only 3.2% and 12.2% of the 2,723 automatically identified image frames required significant manual modifications for lumen and calcified plaque, respectively. Note that results were obtained using images without side branches, since current software does not account for this. In addition, the software missed only 6.7% of frames with calcified plaques and falsely identified only 4.5% of frames with calcifications over all pullbacks. For each pullback, image loading and manual setup took approximately 20 seconds, followed by fully automated processing that took approximately 4 minutes per pullback, or around 0.6 seconds per frame. It is possible to setup multiple pullbacks at once into a queue. After automated analysis, we measured the manual review/editing time per lesion because this is more important metric for user intervention. Only approximately 3.8% of pixels within a lesion were significantly modified, leading to an average editing time of 7.5 seconds per frame, a tremendous reduction (≈80%) compared to complete manual analysis, which requires 36.5 seconds per frame.

## 5   Clinical Research Applications

With OCTOPUS, we can uniquely analyze serial imaging data. We utilized OCTOPUS to analyze serial imaging data from a project on the long-term characterization (18 months) of stented vessels with drug-eluting stents, showing the evolution of neo-atherosclerosis. Neo-atherosclerosis was defined as a lipid neo-intima or a calcified neo-intima [38]. We analyzed pre-stent, 3-month follow-up, and 18-month follow-up IVOCT images to evaluate vessel healing and development of neo-atherosclerosis (Fig. 7). Pre-stent pullback was automatically analyzed using the *Baseline* mode (Fig. 2), and OCTOPUS provided segmentation results and quantitative measurements. Follow-up pullbacks were analyzed using the *Stent Analysis* mode (Fig. 4). If automated results (e.g., plaque segmentation and stent strut detection) were not acceptable, we manually edited/updated results. We then loaded all analyzed pullbacks to the *Follow-up* mode (Fig. 7), registered follow-up pullbacks to pre-stent pullback, and performed further analysis. Using OCTOPUS, we found there was a small region (3.1 *mm*) with uncovered struts at 3-months. In the exact same region, all struts were covered at 18-months. In addition, we observed negative remodeling of the vessel, with the lumen area shrinking from 12.0 $mm^2$ (3-months) to 7.31 $mm^2$ (18-months). The software was used to identify the evolution of neo-atherosclerosis. In Fig. 7, baseline and 3-month follow-up pullbacks showed no clear evidence of lipid. However, significant neo-atherosclerosis formation was detected at the 18-months follow-up of this same region, with a fibrous cap thickness of 0.11 *mm* and an arc angle of 59º containing macrophage infiltration (5 o'clock). We also found neo-intima development with a maximum thickness of about 0.4 *mm* (11-1 o'clock). In Fig. 7, lipid was manually annotated on the 18-month follow-up pullback using our interactive editing tool.

As another example of serial analysis, we determined the role of calcifications in stent deployment using an ex vivo cadaveric coronary artery model (Fig. 8). We assessed stent expansion as a function of balloon size (e.g., nominal, +0.5, +1.0, and +1.5 *mm*) and pressure (e.g., 10, 20, and 30 atm) of non-compliant balloons. In total, we analyzed 14 IVOCT pullbacks with more than 7,500 image frames. OCTOPUS was able to automatically characterize coronary tissues from the baseline and multiple post-stenting pullbacks, co-register pullbacks, and provide quantitative measurements (e.g., lumen area, calcification angle, and calcification thickness). As shown in Fig. 8, we observed that the lumen area gradually increased from 4.12 $mm^2$ to 9.24 $mm^2$ as balloon size and pressure increased. We also compared the length of the malapposed lesion for the entire set of pullbacks. There were 21 *mm* and 10 *mm* of strut malapposition, respectively, for 2.5 *mm* diameter/30 atm pressure and 3.0 *mm* diameter/20 atm pressure conditions. However, the malapposition length was significantly reduced to 2.5 *mm* under the condition of 3.5 *mm* diameter and 30 atm pressure. Comparing the four IVOCT pullbacks, we additionally observed that calcified plaque is fractured based upon varying expansion conditions. Calcium fracture was first observed under the condition of 3.0 *mm* balloon diameter and 20 atm pressure, and the size was gradually expanded. Figure 8 shows an example of ex vivo analysis using OCTOPUS (*Follow-up 4* mode).

## 6   Discussion and Conclusions

We described the development and application of Optical Coherence TOmography PlaqUe and Stent (OCTOPUS) software for quantitative and visual analysis of plaques and stents in IVOCT images of coronary arteries. To the best of our knowledge, this is the first report of a comprehensive software package created to support a variety of clinical research projects. It includes important features and innovations against currently available commercial software, such as plaque characterization, stent deployment analysis (e.g., tissue coverage and malapposition quantification), and registration of pullbacks. The software not only supports single and multiple pullback analysis, but can also be applied to the entire pullback or specific ROIs determined by the user. OCTOPUS also facilitates analysis of plaque morphology through cross-sectional, longitudinal, and en face viewers. The follow-up mode enables loading of up to four pullbacks, allowing more efficient analysis of the baseline and follow-up pullbacks acquired at different time points. The user-friendly editing functions are built into the software to enable modification of automated results. The software is highly automated and fully operated through a convenient GUI.

OCTOPUS software greatly reduces the efforts needed to achieve highly accurate labeling of plaques. Typically, it takes up to 45 minutes for clinicians to manually annotate a heavily calcified pullback (depending on the reader's experience and data). OCTOPUS can provide an automated result for a single pullback within 4 minutes, with manual editing completed within a couple minutes. To analyze 100 IVOCT pullbacks for example, the user can see automated results within 6 hours and complete the entire manual modification process in one or two days. In addition, analysis is likely more consistent when using OCTOPUS as compared to a completely manual method.

Automated analysis time could be significantly improved with algorithm optimization. With faster implementation, the software could help interventional cardiologists more accurately make real-time treatment decisions. This would be of utmost importance in the clinical scenario. Once the software provides vessel anatomy, it would enable a proper PCI planning and selection of appropriate treatment for the lesion (i.e., determining the phenotype of the calcification, the necessity of atherectomy, and predicting difficulties during stent implantation). Thus, though OCTOPUS is designed to support clinical research projects, the underlying algorithms may have important clinical applications that would significantly impact patient care. Although we carefully correct frames for clinical research purposes, most of these corrections would not affect clinical decision making. For the most part, frames containing calcified lesions are reliably identified, and the extent of calcification is segmented. We previously reported that we could successfully identify calcifications with sufficient accuracy to warrant clinical use [16]. We performed a clinical score assessment to determine the potential impact of our method on clinical decision making. All cardiologists unanimously scored the highest (strongly agree), indicating strong agreement that clinical decision making would be the same for automated and manual results. Moreover, we found that our automated results [17] correctly predicted the IVOCT calcium score of Fujino et al. [39], which was proposed to predict the success of stent expansion based on identification of lesions needing plaque modification (e.g., atherectomy).

In summary, we introduced and evaluated the clinical application of a highly automated software package, OCTOPUS, for quantitative plaque analysis in IVOCT images. OCTOPUS gives good plaque segmentation results in a reasonable computation time. We found that our software has the potential to enable faster, more accurate assessment of plaque distribution that can better inform treatment decision making. The software is currently used for offline clinical research purposes by cardiologists, but it could have important applications in real-time treatment planning.

## Acknowledgments

This project was supported by the National Heart, Lung, and Blood Institute through grants NIH R21HL108263, NIH R01HL114406, and NIH R01HL143484. This work was also supported by American Heart Association Grant #20POST35210974/Juhwan Lee/2020. This research was conducted in space renovated using funds from an NIH construction grant (C06 RR12463) awarded to Case Western Reserve University. The content of this report is solely the responsibility of the authors and does not necessarily represent the official views of the National Institutes of Health. The grants were obtained via collaboration between Case Western Reserve University and University Hospitals of Cleveland. This work made use of the High-Performance Computing Resource in the Core Facility for Advanced Research Computing at Case Western Reserve University.

## Disclosures

Dr. Bezerra has received consulting fees from Abbott Vascular.

## References

[1] L.K. Kim, D.N. Feldman, R.V. Swaminathan, R.M. Minutello, J. Chanin, D.C. Yang, M.K. Lee, K. Charitakis, A. Shah, R.K. Kaple, G. Bergman, H. Singh, S.C. Wong, Rate of Percutaneous Coronary Intervention for the Management of Acute Coronary Syndromes and Stable Coronary Artery Disease in the United States (2007 to 2011), The American Journal of Cardiology. 114 (2014) 1003–1010. https://doi.org/10.1016/j.amjcard.2014.07.013.

[2] M.I. Tomey, A.S. Kini, S.K. Sharma, Current Status of Rotational Atherectomy, JACC: Cardiovascular Interventions. 7 (2014) 345–353. https://doi.org/10.1016/j.jcin.2013.12.196.

[3] G.D. Dangas, B.E. Claessen, A. Caixeta, E.A. Sanidas, G.S. Mintz, R. Mehran, In-Stent Restenosis in the Drug-Eluting Stent Era, Journal of the American College of Cardiology. 56 (2010) 1897–1907. https://doi.org/10.1016/j.jacc.2010.07.028.

[4] G.F. Attizzani, D. Capodanno, Y. Ohno, C. Tamburino, Mechanisms, Pathophysiology, and Clinical Aspects of Incomplete Stent Apposition, Journal of the American College of Cardiology. 63 (2014) 1355–1367. https://doi.org/10.1016/j.jacc.2014.01.019.

[5] Lüscher Thomas F., Steffel Jan, Eberli Franz R., Joner Michael, Nakazawa Gaku, Tanner Felix C., Virmani Renu, Drug-Eluting Stent and Coronary Thrombosis, Circulation. 115 (2007) 1051–1058. https://doi.org/10.1161/CIRCULATIONAHA.106.675934.

[6] Pfisterer Matthias E., Late Stent Thrombosis After Drug-Eluting Stent Implantation for Acute Myocardial Infarction, Circulation. 118 (2008) 1117–1119. https://doi.org/10.1161/CIRCULATIONAHA.108.803627.

[7] F. Prati, E. Regar, G.S. Mintz, E. Arbustini, C. Di Mario, I.-K. Jang, T. Akasaka, M. Costa, G. Guagliumi, E. Grube, Y. Ozaki, F. Pinto, P.W.J. Serruys, Expert review document on methodology, terminology, and clinical applications of optical coherence tomography: physical principles, methodology of image acquisition, and clinical application for assessment of coronary arteries and atherosclerosis, Eur Heart J. 31 (2010) 401–415. https://doi.org/10.1093/eurheartj/ehp433.

[8] H.G. Bezerra, M.A. Costa, G. Guagliumi, A.M. Rollins, D.I. Simon, Intracoronary optical coherence tomography: a comprehensive review clinical and research applications, JACC Cardiovasc Interv. 2 (2009) 1035–1046. https://doi.org/10.1016/j.jcin.2009.06.019.

[9] Yabushita Hiroshi, Bouma Brett E., Houser Stuart L., Aretz H. Thomas, Jang Ik-Kyung, Schlendorf Kelly H., Kauffman Christopher R., Shishkov Milen, Kang Dong-Heon, Halpern Elkan F., Tearney Guillermo J., Characterization of Human Atherosclerosis by Optical Coherence Tomography, Circulation. 106 (2002) 1640–1645. https://doi.org/10.1161/01.CIR.0000029927.92825.F6.

[10] I.-K. Jang, G.J. Tearney, B. MacNeill, M. Takano, F. Moselewski, N. Iftima, M. Shishkov, S. Houser, H.T. Aretz, E.F. Halpern, B.E. Bouma, In Vivo Characterization of Coronary Atherosclerotic Plaque by Use of Optical Coherence Tomography, Circulation. 111 (2005) 1551–1555. https://doi.org/10.1161/01.CIR.0000159354.43778.69.

[11] G.J. Tearney, E. Regar, T. Akasaka, T. Adriaenssens, P. Barlis, H.G. Bezerra, B. Bouma, N. Bruining, J. Cho, S. Chowdhary, M.A. Costa, R. de Silva, J. Dijkstra, C. Di Mario, D. Dudek, D. Dudeck, E. Falk, E. Falk, M.D. Feldman, P. Fitzgerald, H.M. Garcia-Garcia, H. Garcia, N. Gonzalo, J.F. Granada, G. Guagliumi, N.R. Holm, Y. Honda, F. Ikeno, M. Kawasaki, J. Kochman, L. Koltowski, T. Kubo, T. Kume, H. Kyono, C.C.S. Lam, G. Lamouche, D.P. Lee, M.B. Leon, A. Maehara, O. Manfrini, G.S. Mintz, K. Mizuno, M. Morel, S. Nadkarni, H. Okura, H. Otake, A. Pietrasik, F. Prati, L. Räber, M.D. Radu, J. Rieber, M. Riga, A. Rollins, M. Rosenberg, V. Sirbu, P.W.J.C. Serruys, K. Shimada, T. Shinke, J. Shite, E. Siegel, S. Sonoda, S. Sonada, M. Suter, S. Takarada, A. Tanaka, M. Terashima, T. Thim, T. Troels, S. Uemura, G.J. Ughi, H.M.M. van Beusekom, A.F.W. van der Steen, G.-A. van Es, G.-A. van Es, G. van Soest, R. Virmani, S. Waxman, N.J. Weissman, G. Weisz, International Working Group for Intravascular Optical Coherence Tomography (IWG-IVOCT), Consensus standards for acquisition, measurement, and reporting of intravascular optical coherence tomography studies: a report from the International Working Group for Intravascular Optical Coherence Tomography Standardization and Validation, J. Am. Coll. Cardiol. 59 (2012) 1058–1072. https://doi.org/10.1016/j.jacc.2011.09.079.

[12] Tearney Guillermo J., Yabushita Hiroshi, Houser Stuart L., Aretz H. Thomas, Jang Ik-Kyung, Schlendorf Kelly H., Kauffman Christopher R., Shishkov Milen, Halpern Elkan F., Bouma Brett E., Quantification of Macrophage Content in Atherosclerotic Plaques by Optical Coherence Tomography, Circulation. 107 (2003) 113–119. https://doi.org/10.1161/01.CIR.0000044384.41037.43.

[13] M. Gargesha, R. Shalev, D. Prabhu, K. Tanaka, A.M. Rollins, M. Costa, H.G. Bezerra, D.L. Wilson, Parameter estimation of atherosclerotic tissue optical properties from three-dimensional intravascular optical coherence tomography, J Med Imaging (Bellingham). 2 (2015) 016001. https://doi.org/10.1117/1.JMI.2.1.016001.

[14] D. Prabhu, H. Bezerra, C. Kolluru, Y. Gharaibeh, E. Mehanna, H. Wu, D. Wilson, Automated A-line coronary plaque classification of intravascular optical coherence tomography images using handcrafted features and large datasets, J Biomed Opt. 24 (2019) 1–15. https://doi.org/10.1117/1.JBO.24.10.106002.

[15] C. Kolluru, D. Prabhu, Y. Gharaibeh, H. Bezerra, G. Guagliumi, D. Wilson, Deep neural networks for A-line-based plaque classification in coronary intravascular optical coherence tomography images, J Med Imaging (Bellingham). 5 (2018) 044504. https://doi.org/10.1117/1.JMI.5.4.044504.

[16] J. Lee, D. Prabhu, C. Kolluru, Y. Gharaibeh, V.N. Zimin, H.G. Bezerra, D.L. Wilson, D.L. Wilson, Automated plaque characterization using deep learning on coronary intravascular optical coherence tomographic images, Biomed. Opt. Express, BOE. 10 (2019) 6497–6515. https://doi.org/10.1364/BOE.10.006497.

[17] Y. Gharaibeh, D.S. Prabhu, C. Kolluru, J. Lee, V. Zimin, H.G. Bezerra, D.L. Wilson, Coronary calcification segmentation in intravascular OCT images using deep learning: application to calcification scoring, JMI. 6 (2019) 045002. https://doi.org/10.1117/1.JMI.6.4.045002.

[18] J. Lee, Y. Gharaibeh, C. Kolluru, V.N. Zimin, L.A.P. Dallan, J.N. Kim, H.G. Bezerra, D.L. Wilson, Segmentation of Coronary Calcified Plaque in Intravascular OCT Images Using a Two-Step Deep Learning Approach, IEEE Access. 8 (2020) 225581–225593. https://doi.org/10.1109/ACCESS.2020.3045285.

[19] C. Kolluru, J. Lee, Y. Gharaibeh, H.G. Bezerra, D.L. Wilson, Learning With Fewer Images via Image Clustering: Application to Intravascular OCT Image Segmentation, IEEE Access. 9 (2021) 37273–37280. https://doi.org/10.1109/ACCESS.2021.3058890.

[20] J. Lee, D. Prabhu, C. Kolluru, Y. Gharaibeh, V.N. Zimin, L.A.P. Dallan, H.G. Bezerra, D.L. Wilson, Fully automated plaque characterization in intravascular OCT images using hybrid convolutional and lumen morphology features, Scientific Reports. 10 (2020) 2596. https://doi.org/10.1038/s41598-020-59315-6.

[21] H. Lu, M. Gargesha, Z. Wang, D. Chamie, G.F. Attizzani, T. Kanaya, S. Ray, M.A. Costa, A.M. Rollins, H.G. Bezerra, D.L. Wilson, Automatic stent detection in intravascular OCT images using bagged decision trees, Biomed Opt Express. 3 (2012) 2809–2824. https://doi.org/10.1364/BOE.3.002809.

[22] H. Lu, J. Lee, S. Ray, K. Tanaka, H.G. Bezerra, A.M. Rollins, D.L. Wilson, Automated stent coverage analysis in intravascular OCT (IVOCT) image volumes using a support vector machine and mesh growing, Biomed. Opt. Express, BOE. 10 (2019) 2809–2828. https://doi.org/10.1364/BOE.10.002809.

[23] H. Lu, J. Lee, M. Jakl, Z. Wang, P. Cervinka, H.G. Bezerra, D.L. Wilson, Application and Evaluation of Highly Automated Software for Comprehensive Stent Analysis in Intravascular Optical Coherence Tomography, Sci Rep. 10 (2020) 2150. https://doi.org/10.1038/s41598-020-59212-y.

[24] Z. Wang, M.W. Jenkins, G.C. Linderman, H.G. Bezerra, Y. Fujino, M.A. Costa, D.L. Wilson, A.M. Rollins, 3-D Stent Detection in Intravascular OCT Using a Bayesian Network and Graph Search, IEEE Trans Med Imaging. 34 (2015) 1549–1561. https://doi.org/10.1109/TMI.2015.2405341.

[25] Z. Wang, H. Kyono, H.G.B. M.d, D.L. Wilson, M.A.C. M.d, A.M. Rollins, Automatic segmentation of intravascular optical coherence tomography images for facilitating quantitative diagnosis of atherosclerosis, in: Optical Coherence Tomography and Coherence Domain Optical Methods in Biomedicine XV, SPIE, 2011: pp. 100–106. https://doi.org/10.1117/12.876003.

[26] Z. Wang, H. Kyono, H.G. Bezerra, H. Wang, M. Gargesha, C. Alraies, C. Xu, J.M. Schmitt, D.L. Wilson, M.A. Costa, A.M. Rollins, Semiautomatic segmentation and quantification of calcified plaques in intracoronary optical coherence tomography images, J Biomed Opt. 15 (2010) 061711. https://doi.org/10.1117/1.3506212.

[27] V. Badrinarayanan, A. Kendall, R. Cipolla, SegNet: A Deep Convolutional Encoder-Decoder Architecture for Image Segmentation, IEEE Transactions on Pattern Analysis and Machine Intelligence. 39 (2017) 2481–2495. https://doi.org/10.1109/TPAMI.2016.2644615.

[28] K. Simonyan, A. Zisserman, Very Deep Convolutional Networks for Large-Scale Image Recognition, ArXiv:1409.1556 [Cs]. (2014). http://arxiv.org/abs/1409.1556 (accessed March 20, 2019).

[29] H. Sinclair, C. Bourantas, A. Bagnall, G.S. Mintz, V. Kunadian, OCT for the Identification of Vulnerable Plaque in Acute Coronary Syndrome, JACC: Cardiovascular Imaging. 8 (2015) 198–209. https://doi.org/10.1016/j.jcmg.2014.12.005.

[30] P. Krähenbühl, V. Koltun, Efficient Inference in Fully Connected CRFs with Gaussian Edge Potentials, in: J. Shawe-Taylor, R.S. Zemel, P.L. Bartlett, F. Pereira, K.Q. Weinberger (Eds.), Advances in Neural Information Processing Systems 24, Curran Associates, Inc., 2011: pp. 109–117. http://papers.nips.cc/paper/4296-efficient-inference-in-fully-connected-crfs-with-gaussian-edge-potentials.pdf (accessed June 16, 2020).

[31] H. Jinnouchi, F. Otsuka, Y. Sato, R.R. Bhoite, A. Sakamoto, S. Torii, K. Yahagi, A. Cornelissen, M. Mori, R. Kawakami, F.D. Kolodgie, R. Virmani, A.V. Finn, Healthy Strut Coverage After Coronary Stent Implantation, Circulation: Cardiovascular Interventions. 13 (2020) e008869. https://doi.org/10.1161/CIRCINTERVENTIONS.119.008869.

[32] M. Joner, A.V. Finn, A. Farb, E.K. Mont, F.D. Kolodgie, E. Ladich, R. Kutys, K. Skorija, H.K. Gold, R. Virmani, Pathology of drug-eluting stents in humans: delayed healing and late thrombotic risk, J. Am. Coll. Cardiol. 48 (2006) 193–202. https://doi.org/10.1016/j.jacc.2006.03.042.

[33] A.T.L. Ong, E.P. McFadden, E. Regar, P.P.T. de Jaegere, R.T. van Domburg, P.W. Serruys, Late angiographic stent thrombosis (LAST) events with drug-eluting stents, J. Am. Coll. Cardiol. 45 (2005) 2088–2092. https://doi.org/10.1016/j.jacc.2005.02.086.

[34] M. Pfisterer, H.P. Brunner-La Rocca, P.T. Buser, P. Rickenbacher, P. Hunziker, C. Mueller, R. Jeger, F. Bader, S. Osswald, C. Kaiser, BASKET-LATE Investigators, Late clinical events after clopidogrel discontinuation may limit the benefit of drug-eluting stents: an observational study of drug-eluting versus bare-metal stents, J. Am. Coll. Cardiol. 48 (2006) 2584–2591. https://doi.org/10.1016/j.jacc.2006.10.026.

[35] G. Guagliumi, M.A. Costa, V. Sirbu, G. Musumeci, H.G. Bezerra, N. Suzuki, A. Matiashvili, N. Lortkipanidze, L. Mihalcsik, A. Trivisonno, O. Valsecchi, G.S. Mintz, O. Dressler, H. Parise, A. Maehara, E. Cristea, A.J. Lansky, R. Mehran, G.W. Stone, Strut coverage and late malapposition with paclitaxel-eluting stents compared with bare metal stents in acute myocardial infarction: optical coherence tomography substudy of the Harmonizing Outcomes with Revascularization and Stents in Acute Myocardial Infarction (HORIZONS-AMI) Trial, Circulation. 123 (2011) 274–281. https://doi.org/10.1161/CIRCULATIONAHA.110.963181.

[36] Y. Gharaibeh, J. Lee, D. Prabhu, P. Dong, V.N. Zimin, L.A. Dallan, H. Bezerra, L. Gu, D. Wilson, Co-registration of pre- and post-stent intravascular OCT images for validation of finite element model simulation of stent expansion, in: Medical Imaging 2020: Biomedical Applications in Molecular, Structural, and Functional Imaging, International Society for Optics and Photonics, 2020: p. 1131717. https://doi.org/10.1117/12.2550212.

[37] Y. Gharaibeh, J. Lee, C. Kolluru, V.N. Zimin, L.A.P. Dallan, H.G. Bezerra, D.L. Wilson, Correction of metallic stent struts and guide wire shadows in intravascular optical coherence tomography images using conditional generative adversarial networks, in: Medical Imaging 2021: Image-Guided Procedures, Robotic Interventions, and Modeling, International Society for Optics and Photonics, 2021: p. 115981U. https://doi.org/10.1117/12.2582177.

[38] D. Nakamura, K. Yasumura, H. Nakamura, Y. Matsuhiro, K. Yasumoto, A. Tanaka, Y. Matsunaga-Lee, M. Yano, M. Yamato, Y. Egami, R. Shutta, Y. Sakata, J. Tanouchi, M. Nishino, Different Neoatherosclerosis Patterns in Drug-Eluting- and Bare-Metal Stent Restenosis — Optical Coherence Tomography Study —, Circulation Journal. 83 (2019) 313–319. https://doi.org/10.1253/circj.CJ-18-0701.

[39] A. Fujino, G.S. Mintz, M. Matsumura, T. Lee, S.-Y. Kim, M. Hoshino, E. Usui, T. Yonetsu, E.S. Haag, R.A. Shlofmitz, T. Kakuta, A. Maehara, A new optical coherence tomography-based calcium scoring system to predict stent underexpansion, EuroIntervention. 13 (2018) e2182–e2189. https://doi.org/10.4244/EIJ-D-17-00962.

**Figures**

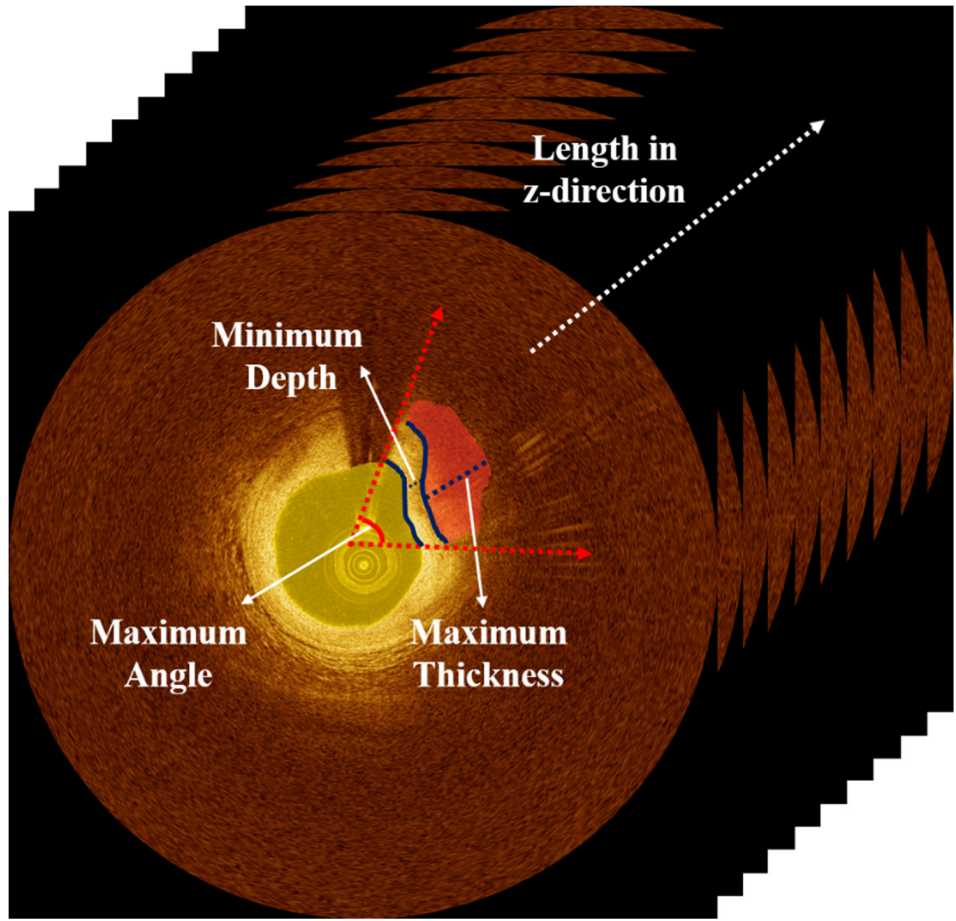

Fig. 1. Quantitative measurement of lumen and calcification attributes. Yellow and red indicate lumen and calcification, respectively.

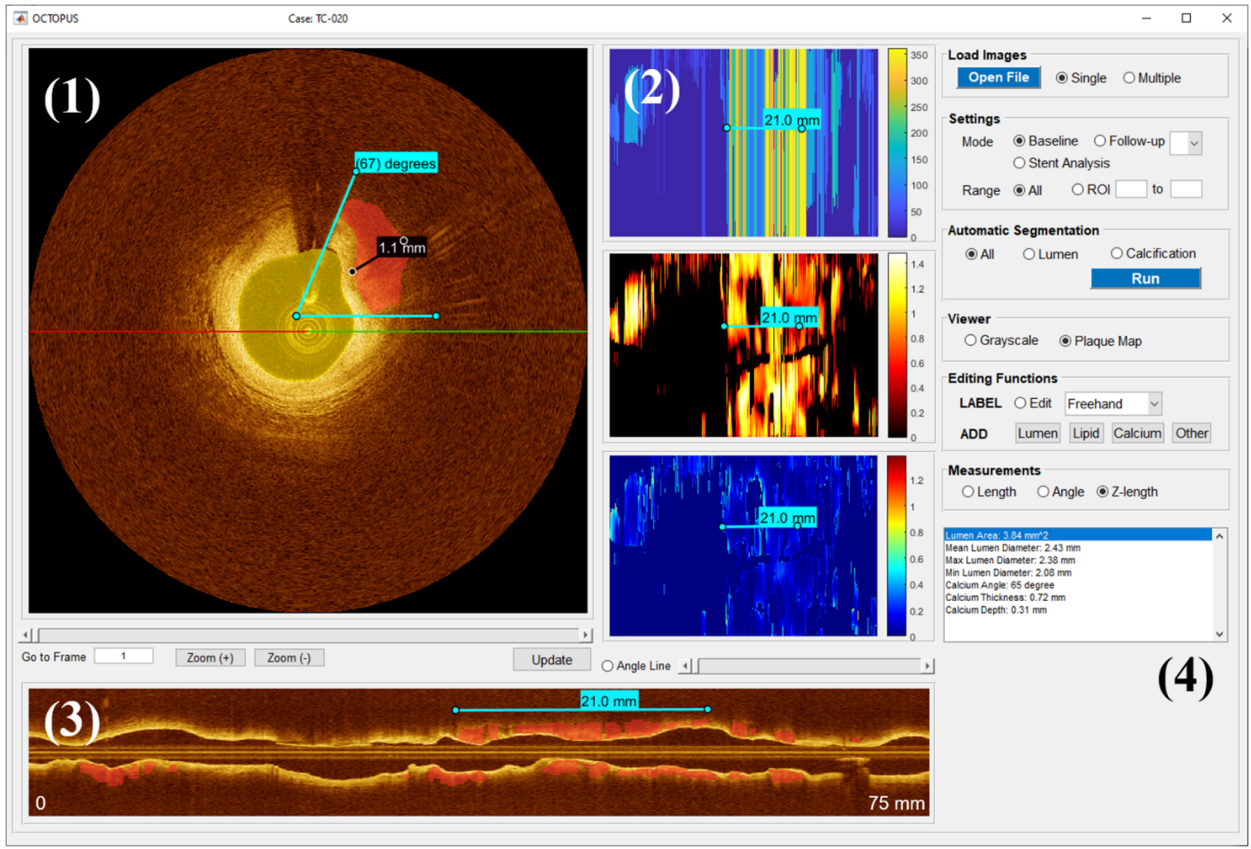

Fig. 2. The GUI of OCTOPUS (*Baseline* mode), divided into four segments: (1) cross-sectional viewer, (2) en face map, (3) longitudinal map, and (4) segmentation functions. The cross-sectional viewer provides frame-wise segmentation results and enables review/update of the automated results. The en face maps represent the unfolded vessel maps as functions of angle (top), thickness (middle), and depth (bottom), and the longitudinal map shows the cut-view of the entire pullback overlaid with calcification labels based on the projection angle (red-green line). Segmentation functions include *Load Images*, *Settings*, *Automatic Segmentation*, *Viewer*, *Editing Functions*, *Measurements*, and a result window.

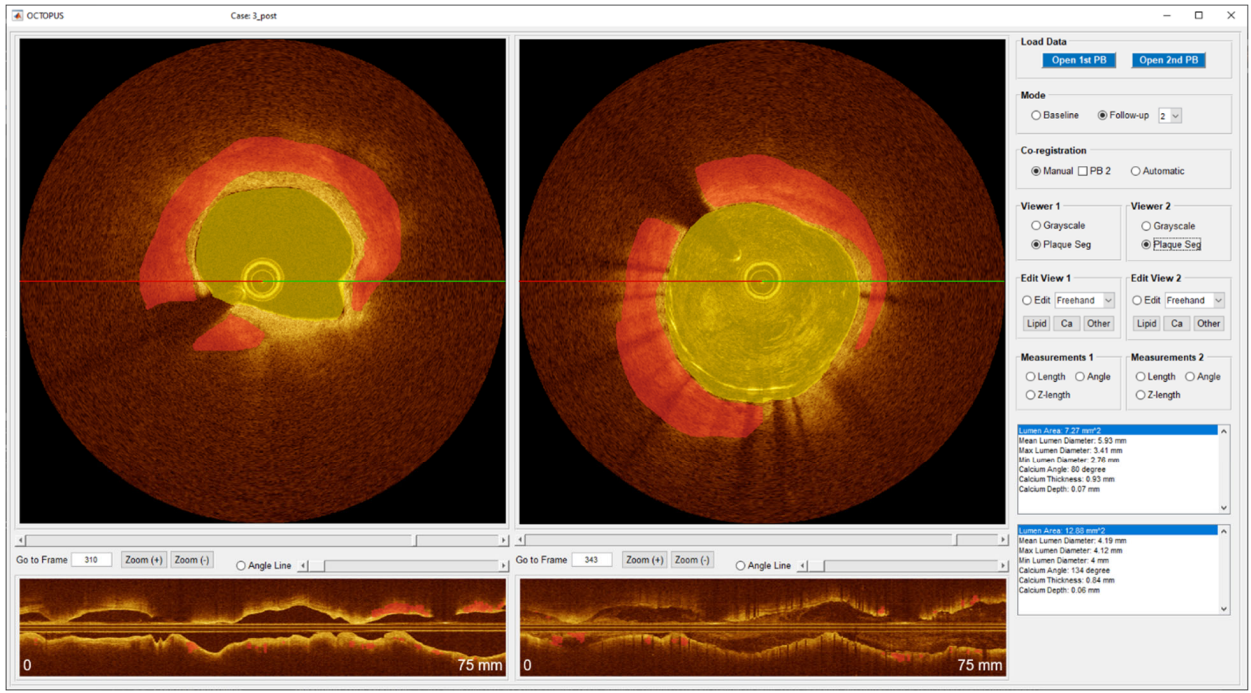

Fig. 3. The GUI of OCTOPUS (*Follow-up 2* mode) consisting of two cross-sectional and two longitudinal viewers and segmentation functions. In this mode, the user can compare two different pullbacks (e.g., baseline and follow-up pullbacks). The left and right IVOCT images are automatically co-registered to baseline and post-stenting pullbacks from the same patient.

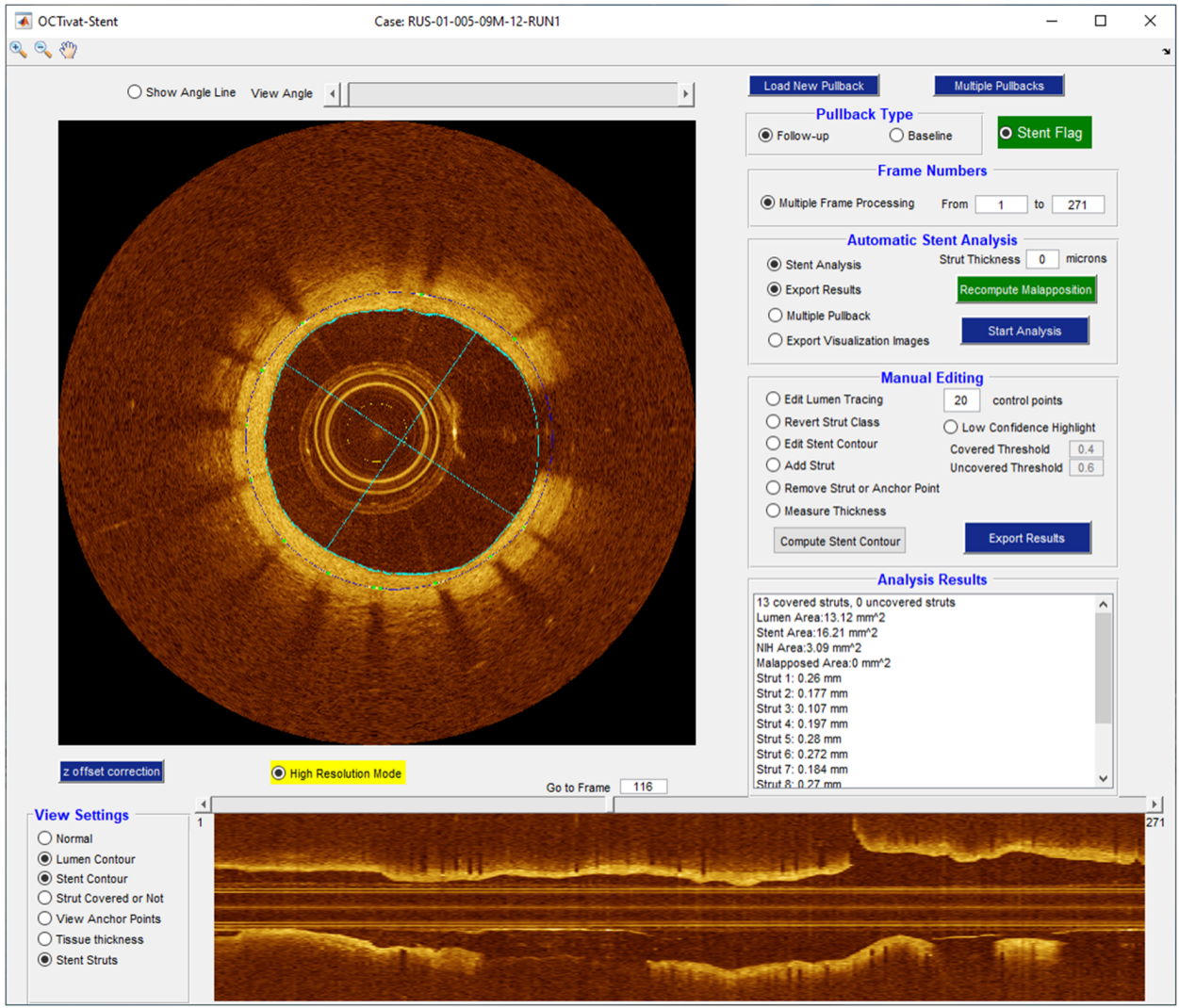

Fig. 4. The GUI of OCTOPUS (*Stent Analysis* mode) enabling comprehensive stent deployment analysis, such as stent strut detection, stent strut classification, stent contour estimation, and tissue coverage and malapposition quantification.

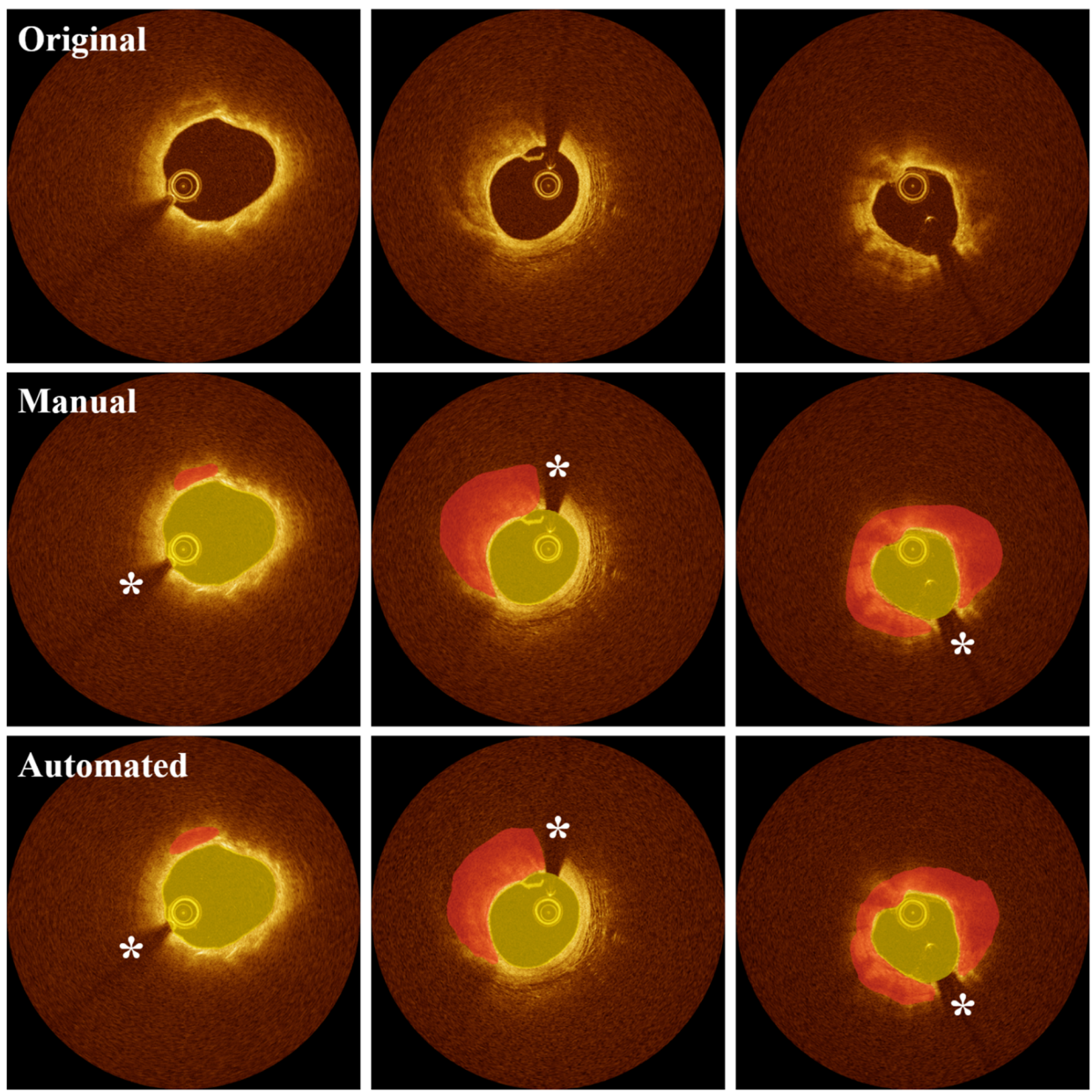

Fig. 5. Representative lumen and calcification segmentation results obtained using OCTOPUS software. (Left) Calcification with angle <90° and maximum thickness <1.0 *mm*, (middle) angle <180° and maximum thickness >1.0 *mm*, and (right) angle <360° and maximum thickness >1.0 *mm*. The top, middle, and bottom rows show original IVOCT images, manual, and automated segmentations, respectively. Manual segmentations (middle) was used to train deep learning models. The software provides reliable segmentation results in most situations. Yellow and red indicate the lumen and calcified plaque, respectively. White asterisk (*) indicates the guidewire shadow.

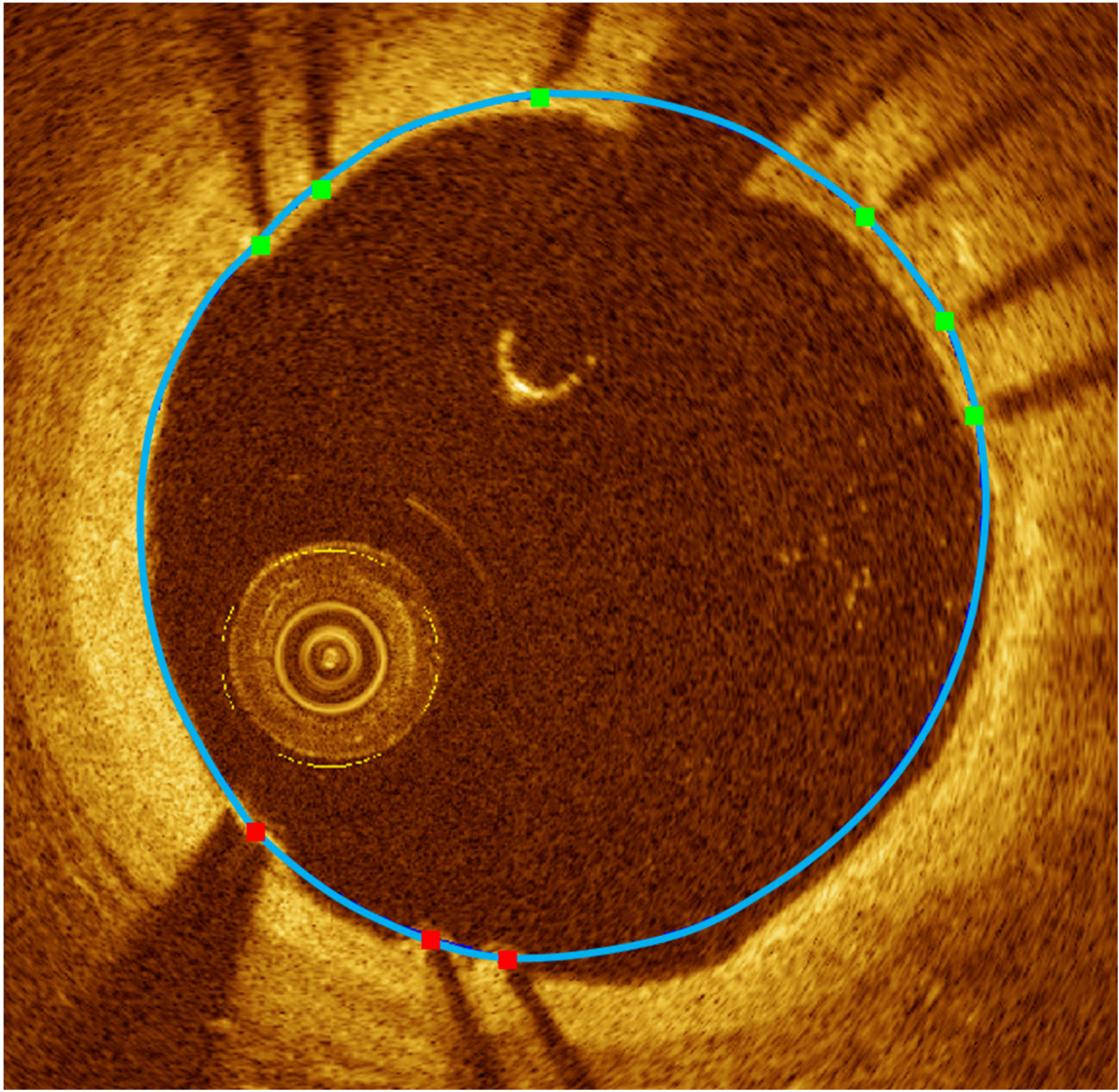

Fig. 6. Example of stent deployment analysis using OCTOPUS. Blue indicates stent contour, green indicates covered struts, and red indicates uncovered struts. The stent contour was created by automatically adding interpolation points between detected struts.

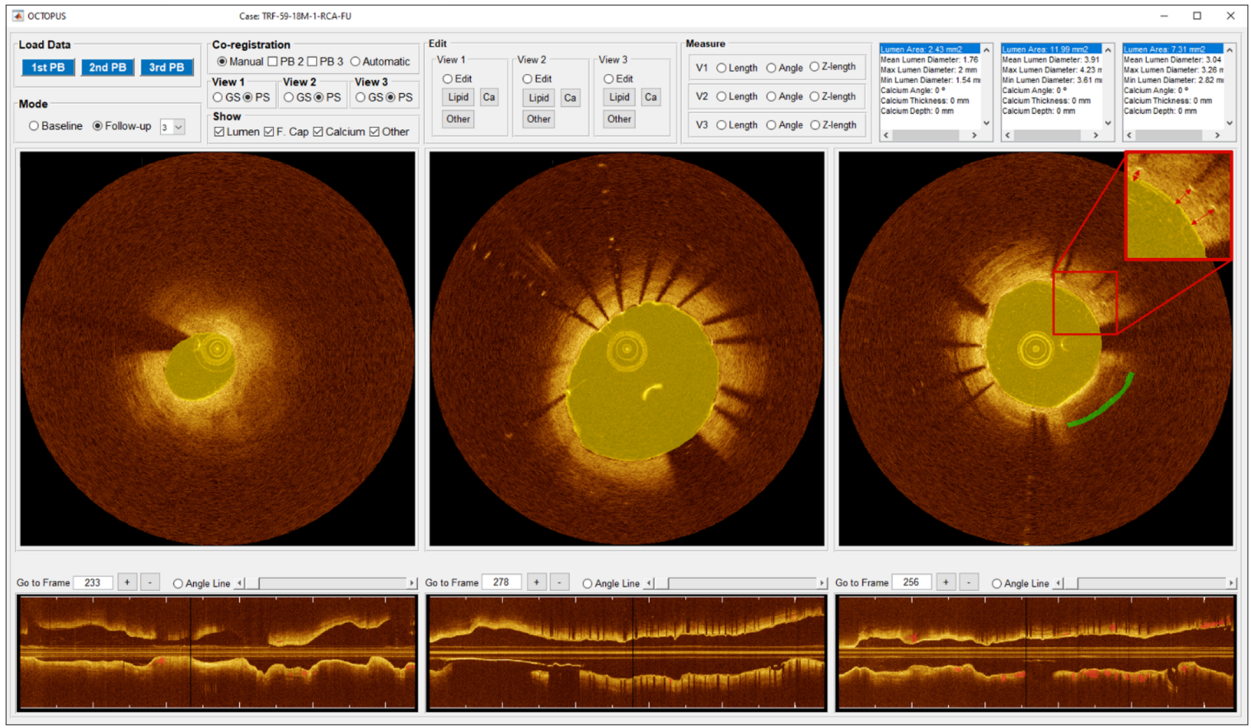

Fig. 7. Analysis of plaque progression/regression in IVOCT images using OCTOPUS (*Follow-up 3* mode). (Left) baseline, (middle) 3-month follow-up, and (right) 18-month follow-up IVOCT pullbacks. Follow-up pullback images were co-registered to the baseline pullback, and plaque characterization was performed. Lipid (green) was manually labeled on the 18-month follow-up pullback using an interactive editing function. Green indicates the development of lipidic neo-atherosclerosis. Red arrow indicates the formation of neo-intima.

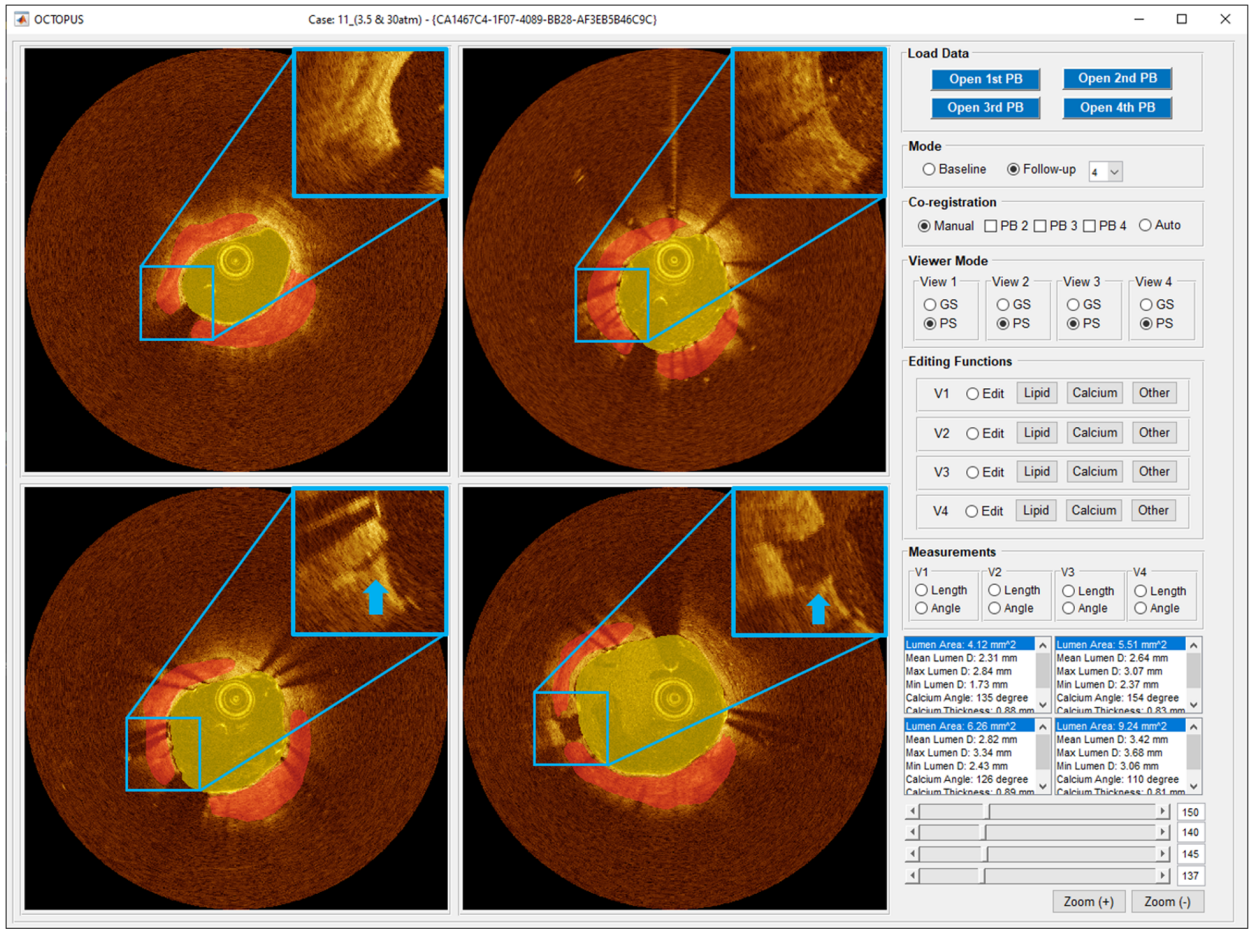

Fig. 8. Analysis of stent expansion in an ex vivo cadaveric study using OCTOPUS (*Follow-up 4* mode). (Top left) baseline, (top right) post-dilatation with 2.5 *mm* diameter and 30 atm pressure, (bottom left) post-dilatation with 3.0 *mm* balloon diameter and 20 atm pressure, and (bottom right) post-dilatation with 3.5 *mm* diameter and 30 atm pressure). Yellow and red indicate lumen and calcium, respectively. Blue arrow indicates calcium fracture.